\documentclass[a4paper,10pt,twoside]{cpc-hepnp}

\usepackage{multicol}
\usepackage{graphicx}
\usepackage{booktabs}
\usepackage{amssymb,bm,mathrsfs,bbm,amscd}
\usepackage[tbtags]{amsmath}
\usepackage{lastpage}
\usepackage{CJK}
\usepackage{subcaption}
\usepackage{hepunits}
\usepackage{hyperref}
\usepackage{lineno}

\usepackage[final]{changes}

\begin{document}
\begin{CJK*}{UTF8}{gbsn}



\fancyhead[c]{\small Submitted to Chinese Physics C} 
\fancyfoot[C]{\small \thepage}
\footnotetext[0]{ }

\title{Development of a two-dimensional imaging GEM detector using the resistive anode readout method with $6\times6$ cells
\thanks{Supported by National Natural Science Foundation of China (11375219)}
\thanks{Supported by the CAS Center for Excellence in Particle Physics (CCEPP)}
}

\author{%
      Xu-Dong Ju(鞠旭东)$^{1,2;1)}$\email{juxd@ihep.ac.cn}%
\quad Ming-Yi Dong(董明义)$^{1,2;2)}$\email{dongmy@ihep.ac.cn} 
\quad Chuan-Xing Zhou(周传兴)$^{1,2,3}$ 
\quad Jing Dong(董静)$^{1,2}$ \\
\quad Yu-Bin Zhao(赵豫斌)$^{1,2}$
\quad Hong-Yu Zhang(章红宇)$^{1,2}$
\quad Hui-Rong Qi(祁辉荣)$^{1,2}$
\quad Qun Ou-Yang(欧阳群)$^{1,2}$
}
\maketitle

\address{%
$^1$ State Key Laboratory of Particle Detection and Electronics, Beijing 100049, China\\
$^2$ Institute of High Energy Physics, Chinese Academy of Sciences, Beijing 100049, China\\
$^3$ University of Chinese Academy of Sciences, Beijing 100049, China\\
}

\begin{abstract}
We report the application of the resistive anode readout method on a two dimensional imaging GEM detector. The resistive anode consists of $6\times6$ cells with the cell size $6~\mathrm{mm}\times6~\mathrm{mm}$. New electronics and DAQ system are used to process the signals from the 49 readout channels. The detector has been tested by using the X-ray tube (8~keV). The spatial resolution of the detector is about $103.46~\mathrm{{\mu}m}$ with the detector response part about $66.41~\mathrm{{\mu}m}$. The nonlinearity of the detector is less than $1.5\%$. A quite good two dimensional imaging capability is achieved as well. The performances of the detector show the prospect of the resistive anode readout method for the large readout area imaging detectors. 

\end{abstract}

\begin{keyword}
resistive anode, GEM, two dimensional, imaging detector, spatial resolution
\end{keyword}

\begin{pacs}
29.40.Cs, 29.40.Gx
\end{pacs}

\footnotetext[0]{\hspace*{-3mm}\raisebox{0.3ex}{$\scriptstyle\copyright$}2013
Chinese Physical Society and the Institute of High Energy Physics
of the Chinese Academy of Sciences and the Institute
of Modern Physics of the Chinese Academy of Sciences and IOP Publishing Ltd}%

\begin{multicols}{2}

\section{Introduction}
Compared with the traditional gaseous detectors like the wire-chamber type detectors, the new type Micro-Pattern Gaseous Detectors (MPGD)~\cite{Shekhtman-2002} like the Gas Electron Multipliers (GEMs)~\cite{Sauli-1997}, have the most outstanding advantages in the spatial resolution as well as the counting rates. However, the number of the detector readout electronics increases rapidly with the high spatial resolution which is closely related to the pitch between two adjacent readout structures. For example, the readout strip pitch of the GEM detectors used in the CERN-COMPASS experiment~\cite{Altunbas-2002,Thibaud-2014} is $400~\mathrm{{\mu}m}$ for a $80~\mathrm{{\mu}m}$ spatial resolution ($\sigma$). The small readout pitch leads to the high density readout electronics. 
\par\indent
To overcome the problem, ASIC readout electronics are developed for the MPGD by the electronics researchers, such as the APV25 chip used in the COMPASS experiment~\cite{Altunbas-2002}. For the detector researchers, meanwhile, a lot of readout methods are being studied to save the electronics.        
\par\indent
The resistive anode readout method is one of the traditional readout methods~\cite{Doke-1987}, which is widely used in the Position Sensitive Silicon Detectors (PSSD)~\cite{Banu-2008} and the Micro-Channel Plates (MCP)~\cite{Lampton-1979}. According to this four-corner resistive readout concept, Sarvestani et al.~\cite{Sarvestani-1998} developed a 2-D interpolating resistive readout structure and applied it for the Micro-CAT detector~\cite{Orthen-2002,Wagner-2002,Wagner-2007}. This kind of resistive pad array structure helps to obtain a good spatial resolution with less electronics. 
\par\indent
We have developed a series of triple-GEM detectors to study this kind of readout method as well, and good detector performances have been achieved~\cite{Dongmy-2013,Xiuql-2013,Juxd-2016}. However, in our previous prototype studies~\cite{Juxd-2016}, only the basic reconstruction element of this readout structure, which consists of $3\times3$ cells, is used availably as the readout anode of the detector, because of the limitation of the electronics. 
\par\indent
In this paper, we try to apply the resistive anode readout method for a relatively larger sensitive readout area with $6\times6$ cells by improving the electronics and DAQ system of the GEM detector. The main purpose of the research is to inspect the two dimensional imaging performance of this readout method for a large area GEM detector. Parameters that influence the imaging quality like the spatial resolution and the nonlinearity are tested. At the same time, the cell size of the resistive anode is chosen to be $\mathrm{6~mm\times6~mm}$ instead of $\mathrm{8~mm\times8~mm}$~\cite{Dongmy-2013} in order to study the spatial resolution capability.
\section{Detector setup}
\subsection{The resistive anode readout board}
The schematic of the elementary cell of the resistive anode readout structure is shown in Fig.~\ref{fig_Anode_Shematics}. Instead of being collected by the readout electrode immediately, charges will diffuse on the high resistive pad (the large black square in Fig.~\ref{fig_Anode_Shematics}), be obstructed by the low resistive strips (the narrow pink lines in Fig.~\ref{fig_Anode_Shematics}) and finally be collected by the readout nodes (the small grey squares in Fig.~\ref{fig_Anode_Shematics}). According to the optimisation of the resistive anode mentioned in Ref.~\cite{Juxd-2016}, the surface resistivity of the pad and the strip is set to be $150~\mathrm{k\Omega/\square}$ and $1~\mathrm{k\Omega/\square}$, respectively. 
\begin{center}
    \includegraphics[width=5cm]{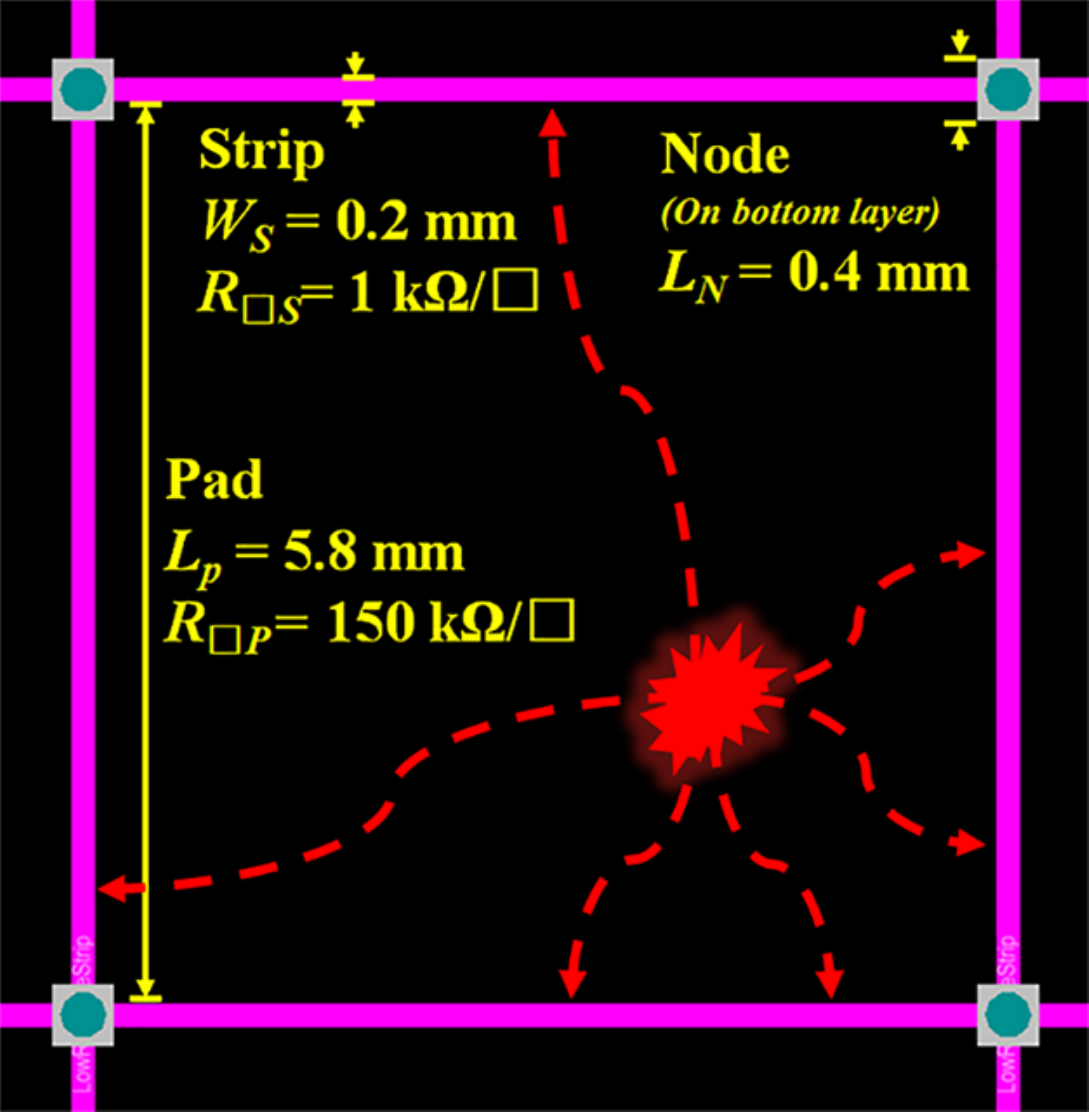}
    \figcaption{\label{fig_Anode_Shematics} (color online) Schematic of the elementary cell of the resistive anode readout board, which includes the high resistive pad (the large black square), the low resistive strips (the narrow pink lines) and the readout nodes (the small grey squares). The yellow words are the design parameters of the resistive anode used in this experiment.}
\end{center}
\par\indent
Fig.~\ref{fig_Anode_Front} shows the overall design of the resistive anode readout board used in the experiment. For this board, there are $6\times6$ cells with the cell size $\mathrm{6~mm\times6~mm}$, which means that the board covers $\mathrm{36\times36~mm^2}$ sensitive area with 49 readout channels. Based on the thick-film resistor process, the board is manufactured by brushing  the resistance slurry onto the 1-mm-thick ceramic board, and then being sintered in the oven with about 825~$^{\circ}$C temperature. 
\begin{center}
    \includegraphics[width=8cm]{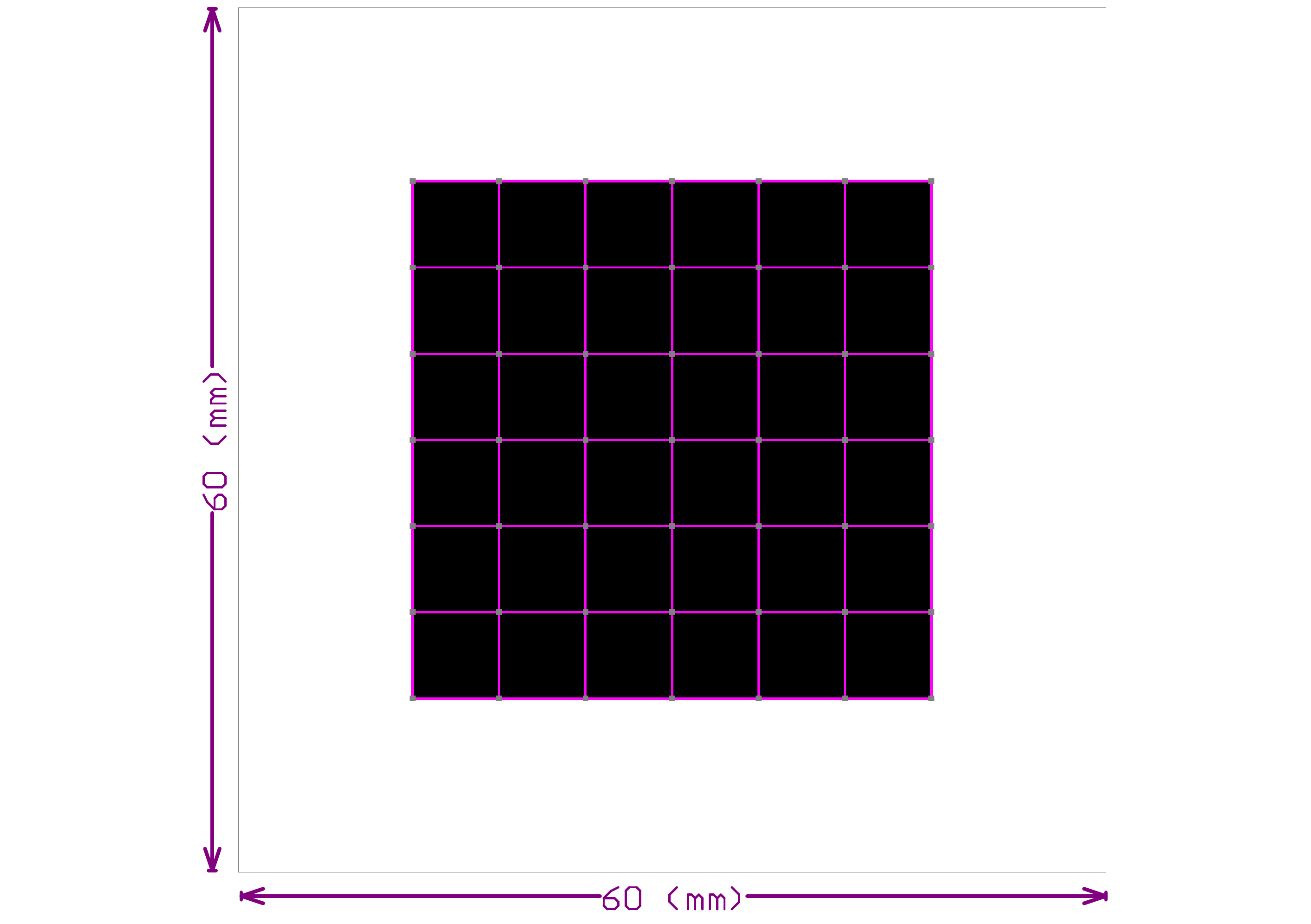}
    \figcaption{\label{fig_Anode_Front} (color online) Resistive layer design of the resistive anode readout board. The large black squares are the high resistive pads. The narrow pink lines are the low resistive strips. The small grey squares are the vias that connect the top resistive layer with the bottom routing layer (not shown in this picture). More details can be seen by zooming out the picture.}
\end{center}
\subsection{The electronics and DAQ}
As shown in Fig.~\ref{fig_Detector_Schematics}, the basic structure of the detector is similar to what has been presented in Ref.~\cite{Juxd-2016}. Three cascading standard GEM foils from CERN work as the gas ($\mathrm{Ar/CO_2(70/30)}$) gain device. Electrons generated in the drift region by incoming particles are multiplied in the holes of the GEM foils and collected by the readout nodes on the resistive anode finally. New readout electronics and DAQ system are used in the experiment.
\begin{center}
    \includegraphics[width=8cm]{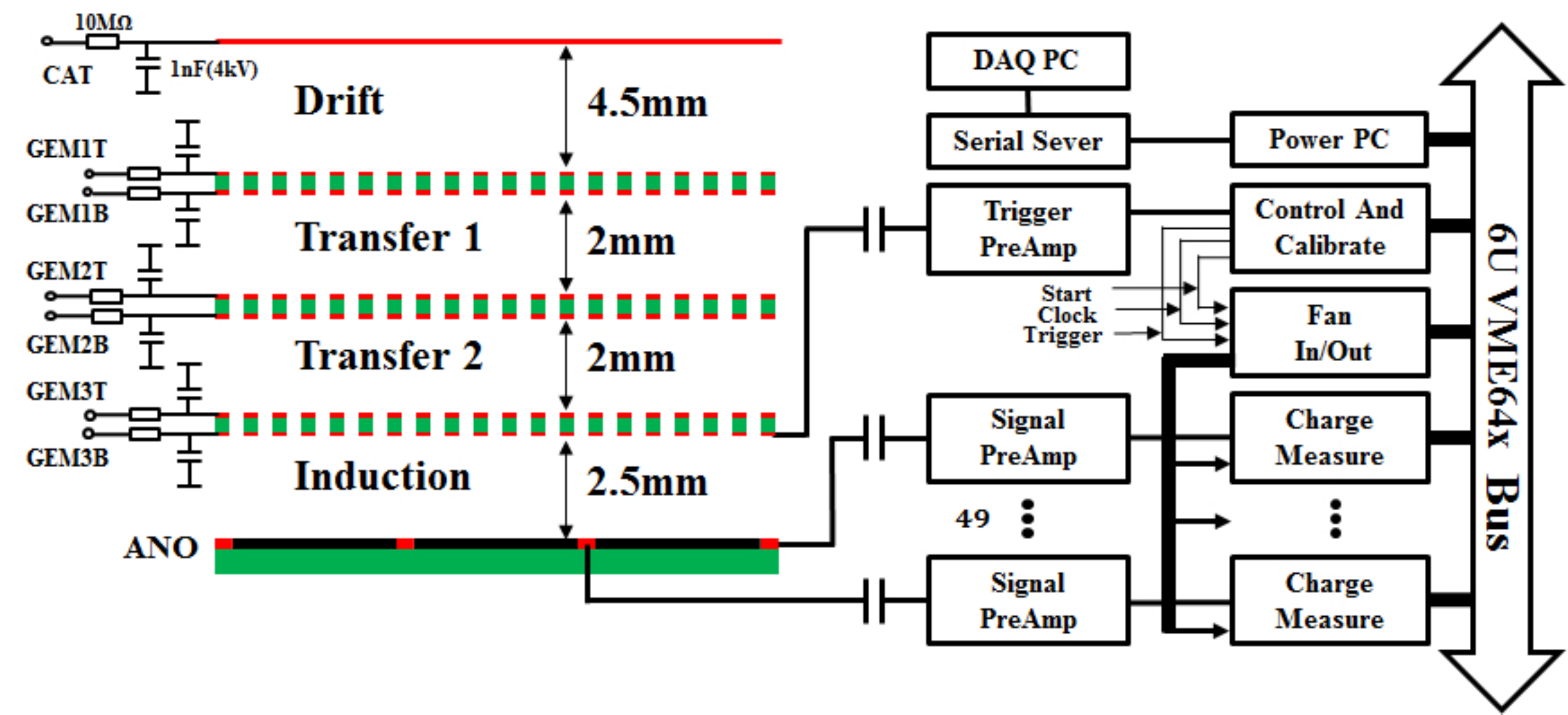}
    \figcaption{\label{fig_Detector_Schematics} (color online) Schematic of the detector and the electronics. Eight signal preamps are used in the experiment and 49 of the total 64 readout channels are active.}
\end{center}
\par\indent
The signal getting from the readout node of the resistive anode is amplified by the Signal PreAmp (charge sensitive preamplifier) and sent to the Charge Measure (CM) module for the following processing. The signal will be amplified further, filtered, shaped, and digitized by a 10-bit ADC. Finally, the peek value of the signal will be seeked by the CM module. 
\par\indent
Different from the previous DAQ system~\cite{Juxd-2016}, the signal getting from the bottom layer of the third GEM foil (GEM3B) works as the trigger signal~\cite{Guedes-2003}. The trigger signal is amplified by the Trigger PreAmp (charge sensitive preamplifier) and sent to the Control And Calibration (CAC) module. The CAC module processes the trigger signal and output the trigger signal level (LVPECL level) together with the start and clock signal levels to the Fan In/Out (FIO) module to control the data acquisition. Fig.~\ref{fig_Oscilloscope} shows the trigger signal (yellow line with mark 1) getting from the oscilloscope, and the green line with mark 2 is the corresponding trigger signal level. There is a strict one-to-one correspondence between the trigger signal and the channel signal (the red line with mark 4).
\begin{center}
    \includegraphics[width=8cm]{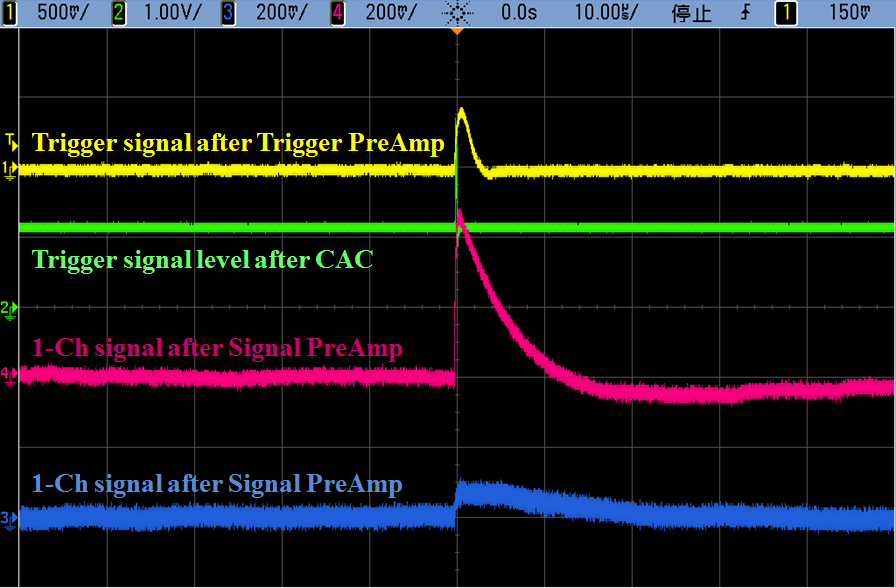}
    \figcaption{\label{fig_Oscilloscope} (color online) Trigger signal from GEM3B after the trigger PreAmp (a charge sensitive amplifier). The picture is getting from the oscilloscope. The yellow line with mark 1 is the trigger signal after the trigger PreAmp. The green line with mark 2 is the trigger signal level which is the output of the CAC module. The blue line with mark 3 and the red line with mark 4 are signals of two different channles after the signal PreAmp.}
\end{center}
\par\indent
The FIO module distributes the trigger and control signal levels to each CM module. All signals are transfered by the VME-Bus. The DAQ system includes programs for the host-computer (DAQ PC) and the slave-computer (Power PC). The data communication between the DAQ PC and the Power PC is through a web server.
\subsection{The electronics calibration and baseline} 
The CAC module takes charge of the calibration of the electronics by an automatical scan of the given DAC which represent the output of the detector. Fig.~\ref{fig_Electronics_Calibration}-a shows the linear relationships of the 49 channels between the DAC from the CAC module and the ADC from the CM module, and the mean value of the slopes in Fig.~\ref{fig_Electronics_Calibration}-a is $3.139\pm0.109$ as shown in Fig.~\ref{fig_Electronics_Calibration}-b.
\begin{center}
    \includegraphics[width=8cm]{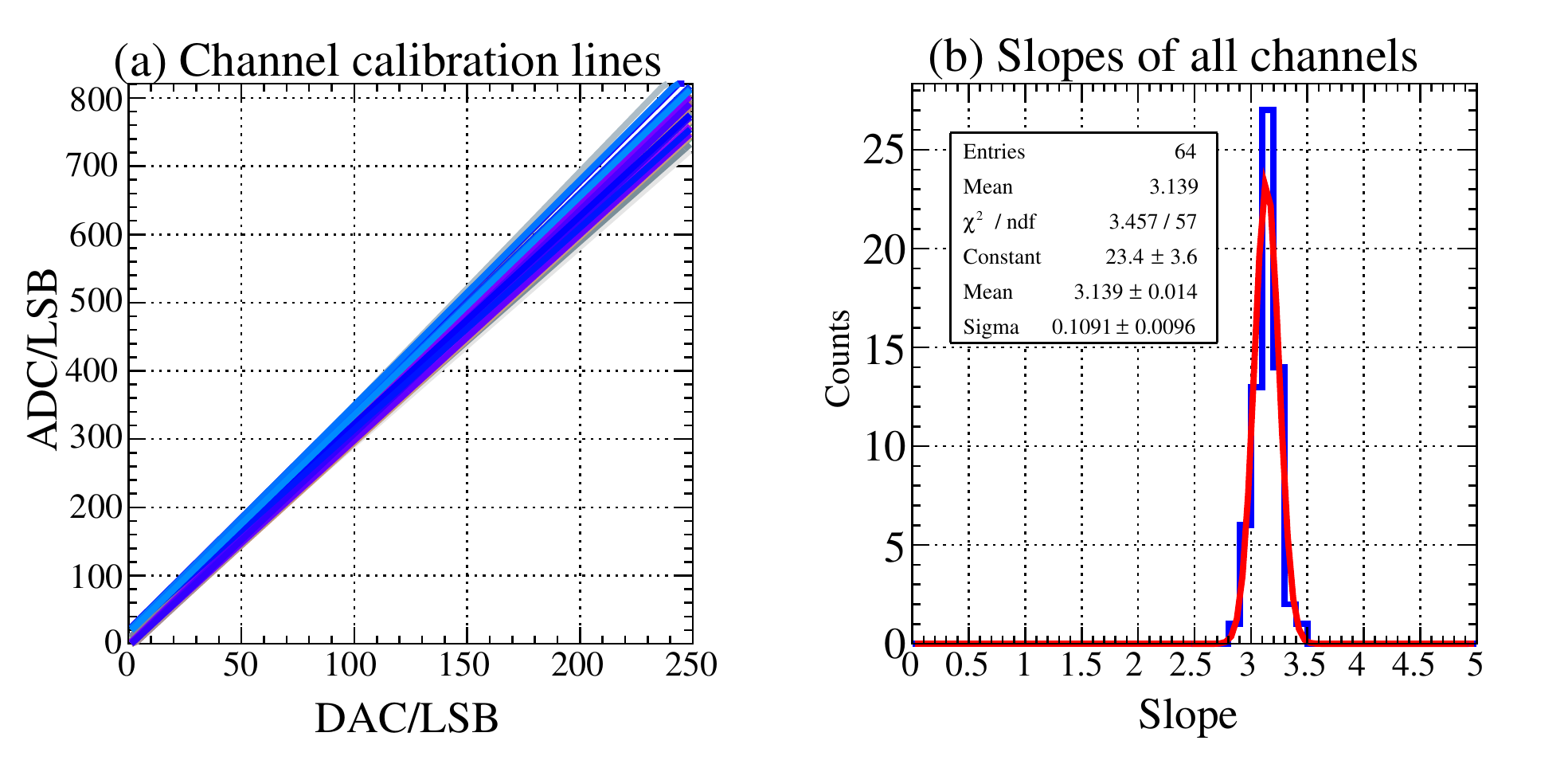}
    \figcaption{\label{fig_Electronics_Calibration} (color online) Calibration of the 49 channels. (a) Linear relationships of the 49 channels between the input DAC of the CAC module and the output ADC from the CM module. (b) Slopes statistics of the 49 channels by fitting with a gaussian distribution.}
\end{center}
\par\indent
The baseline is obtained by the data acquisition when the threshold is set to be 0. The mean value of the gaussian fit of the data is chosen to be the baseline of each channel as shown in Fig.~\ref{fig_Electronics_Baseline}-a. The mean baseline of all 49 channels is $33.17\pm2.07~LSB$ as shown in Fig.~\ref{fig_Electronics_Baseline}-b.
\begin{center}
    \includegraphics[width=8cm]{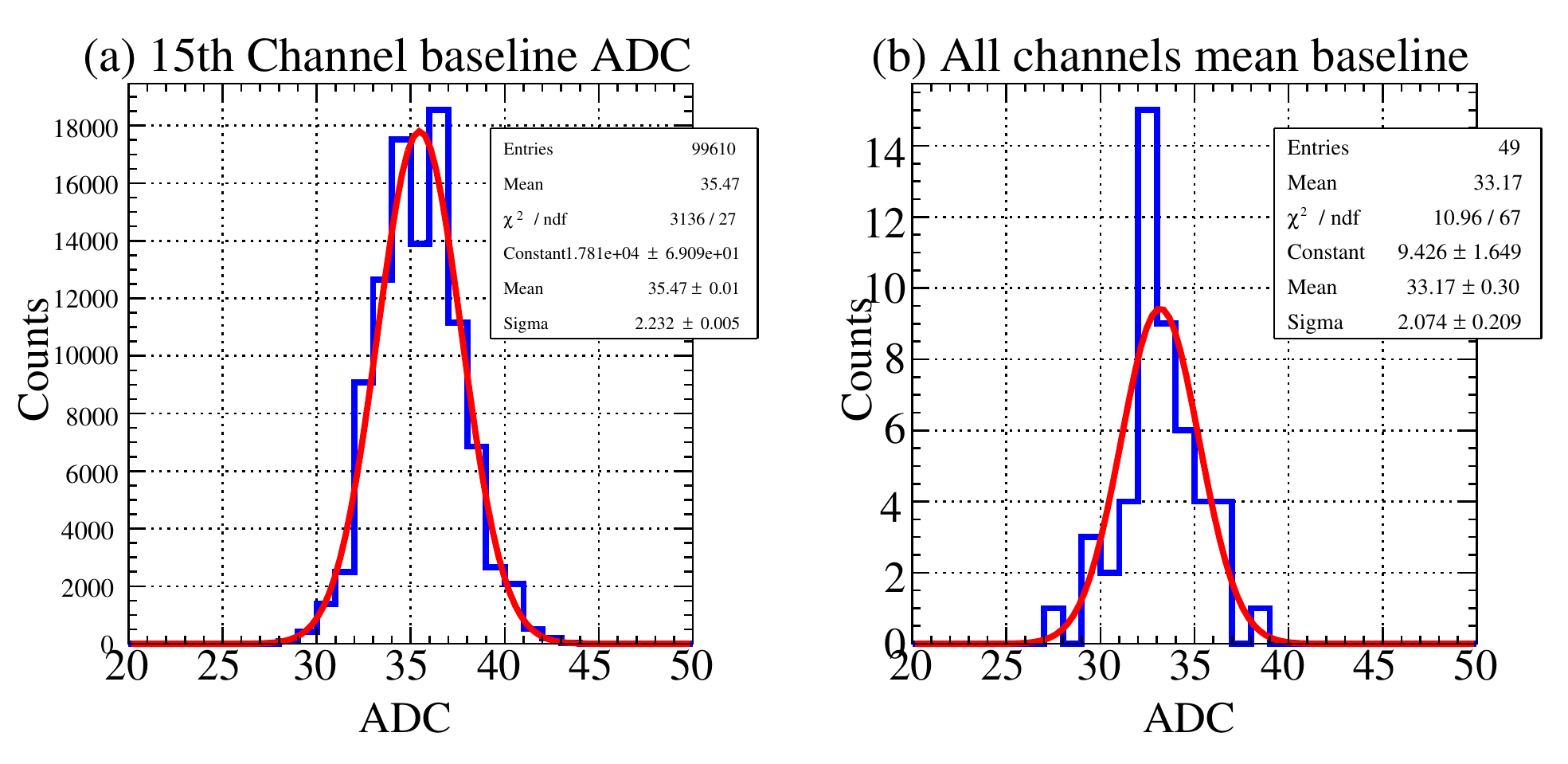}
    \figcaption{\label{fig_Electronics_Baseline} (color online) Baseline of the 49 channels. (a) Baseline of the 15th channel fitted by a gaussian distribution. (b) Baselines (gaussian mean values in (a)) of the 49 channels by fitting with a gaussian distribution.}
\end{center}
\section{Detector performance}
In this study, we pay more attention to the two dimensional imaging performence of the GEM detector using a relatively larger resistive anode. Instead of the ${}^{55}$Fe source (5.9~keV), the X-ray tube (8~keV) is used to test the detector, and the setup of the test is shown in Fig.~\ref{fig_Detector_Test}. 
\begin{center}
    \includegraphics[width=8cm]{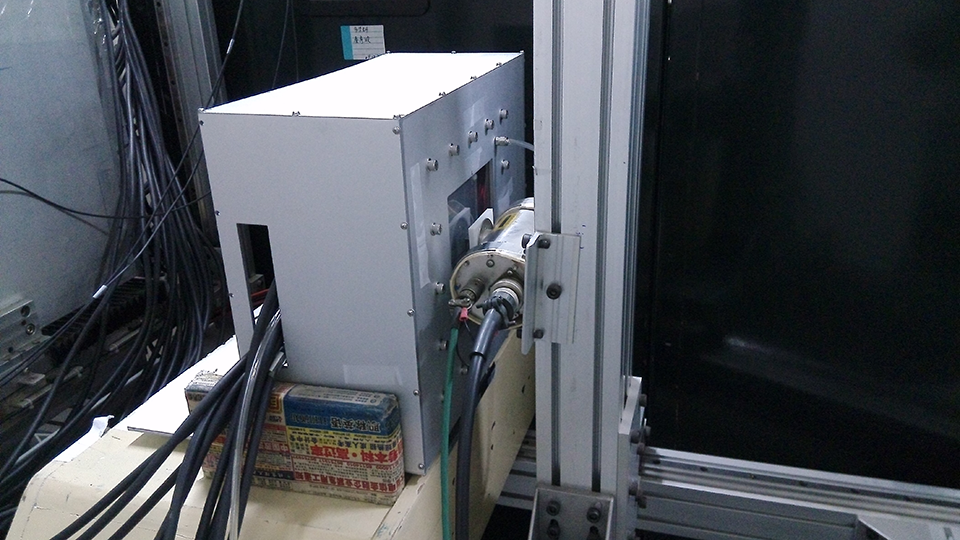}
    \figcaption{\label{fig_Detector_Test} (color online) Test setup of the detector for the spatial resolution.}
\end{center}
\subsection{The spatial resolution}
The spatial resolution is one of the most important factors that reflects the imaging capability of the detectors. Due to the limitation of the beam time, some convenient methods~\cite{Lvxy-2012}, such as the narrow slot imaging and the blade edge imaging, can be used to study the spatial resolution, instead of the conventional beam test method. The cardinal principle of such methods is that the measurement distribution of the ray getting from the detector ($M(x')$) can be treated as the convolution of the true distribution of the ray hitting on the surface of the detector ($T(x)$) and the detector resolution function ($R(x, x')$) like~\cite{Zhuys-2006}
\begin{equation}
M(x')=T(x) \otimes R(x, x')=\int_{-\infty}^{\infty} T(x)R(x, x')dx
\label{Eq_SR_Principle}
\end{equation}
Based on the central limit theorem~\cite{Zhuys-2006}, the detector resolution ($R(x, x')$), which is also the measurement error, obeys the standard normal distribution like 
\begin{equation}
R(x,x')=\frac{1}{\sqrt{2\pi}\sigma}\exp\left[ - \frac{(x-x')^2}{2{\sigma}^2}\right] 
\end{equation}
where $\sigma$ represents the intrinsic spatial resolution of the detector.
\par\indent
Supposing that $T(x)$ is already known, $R(x, x')$ can be extracted by the fitting of $M(x')$. For example, if $T(x)=\delta(x-x_0)$, 
\begin{equation}
M(x')=\int_{-\infty}^{\infty} R(x,x')\delta(x-x_0)dx=R(x_0,x')
\label{Eq_SR_Principle_Detlta}
\end{equation}
which means that $M(x')$ is the same as $R(x_0, x')$. In other words, $R(x_0, x')$ can be easily obtained by using the imaging of an infinite narrow slot. In the experiment, the imaging of a very thin slot is used to study $R(x, x')$ approximately.
\par\indent
Fig.~\ref{fig_Imaging_Slot} shows the imaging of a 40-${\mu}$m-width steel slot by using the X-ray tube (8~keV). The slot image covers more than four cells. In order to study the uniformity of different cells of the resistive anode, the slot image is divided into four parts and marked as shown in Fig.~\ref{fig_Imaging_Slot}. The dashed pink squares represent the relative position and the size of four cells along y axis. 
\begin{center}
    \includegraphics[width=8cm]{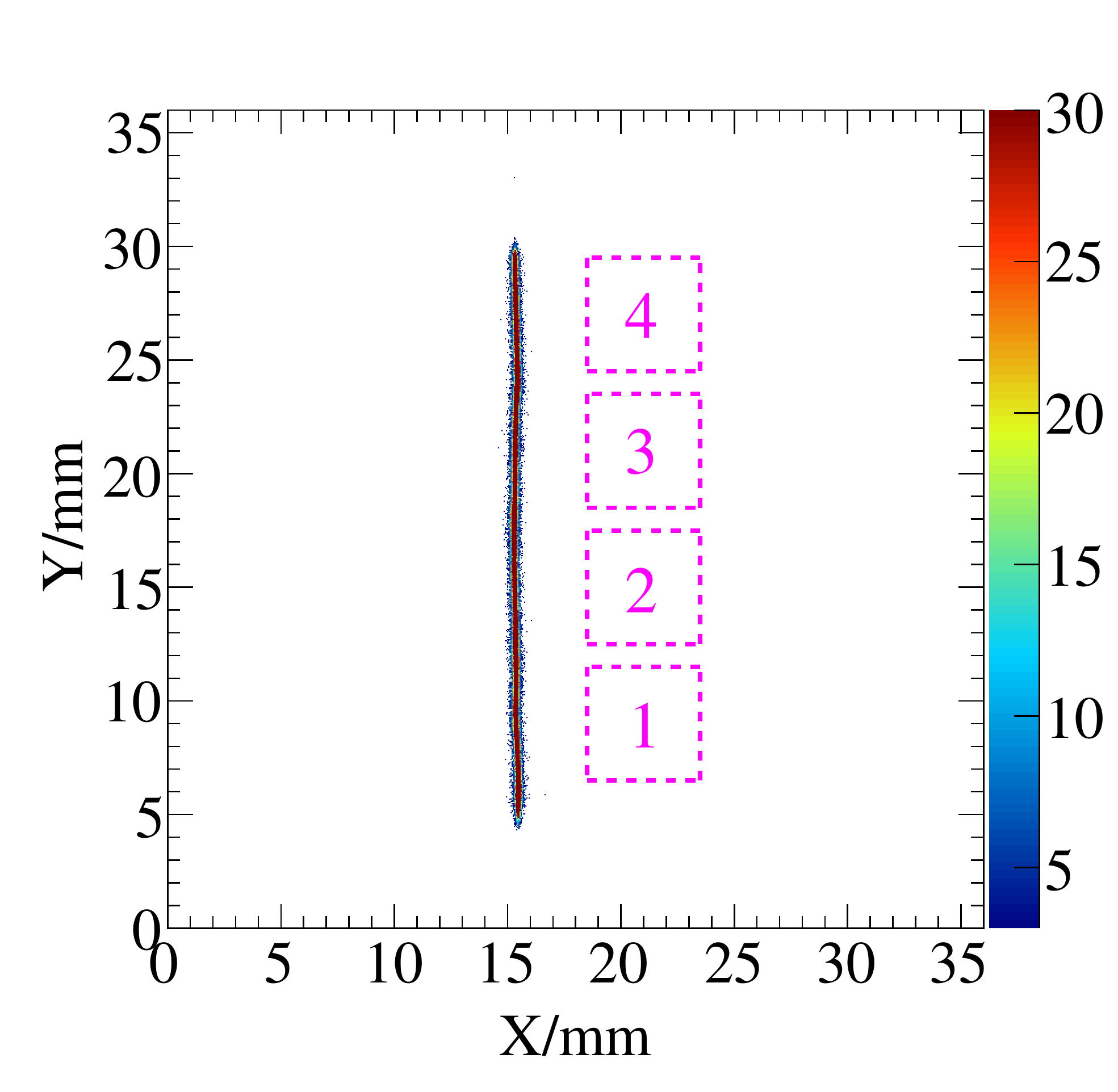}
    \figcaption{\label{fig_Imaging_Slot} (color online) Imaging of the 40-${\mu}$m-width steel slot by using the X-ray tube (8~keV). (See in Fig.~\ref{fig_Detector_Test}) The dashed pink squares are marks of the relative position of four cells along y axis.}
\end{center}
\par\indent
In the experiment, a thin slot is used to approximate the $\delta(x-x_0)$ distribution in Eq.~\ref{Eq_SR_Principle_Detlta}. Because of the slot width effect, it is hard to fit by using a single gaussian distribution. In fact, a composite 2-gaussian model~\cite{Roofit-2006} is used to fit the counts distribution like 
\begin{equation}
\begin{split}
g_D(x) &= N({\mu}_D,{\sigma}_D)  \\
g_S(x) &= N({\mu}_S,{\sigma}_S)  \\
f(x) &= a \cdot g_D(x) + (1-a) \cdot g_S(x) 
\end{split}
\end{equation}
where $g_D(x)$ is the P.D.F of the part that represents the detector response with the mean value ${\mu}_D$ and the spatial resolution $\sigma_D$, $g_B(x)$ is the P.D.F of the part that represents the  effect of the slot width with the mean value ${\mu}_S$ and the spatial resolution $\sigma_S$, $f(x)$ is the composite P.D.F and $a$ is the fraction of the detector response part.
\par\indent
When ${\mu}_D = {\mu}_S$, the spatial resolution ($\sigma$) can be calculated as~\cite{Wulh-2007,Juxd-PhD-2016}
\begin{equation}
\begin{split}
{\sigma}^2 &= \int (x-\mu)^2f(x)dx \\
&=a \cdot \int (x-{\mu}_D)^2g_D(x)dx + (1-a) \cdot \int (x-{\mu}_S)^2g_S(x)dx \\
&=a \cdot {\sigma}_D^2 + (1-a) \cdot {\sigma}_S^2
\end{split}
\end{equation}
\par\indent
Fig.~\ref{fig_SR_Slot-40_Mark2} shows the fitting result of the mark 2 part data in Fig.~\ref{fig_Imaging_Slot}. The spatial resolution obtained by using the 40-${\mu}$m-width slot is about $103.46~\mathrm{{\mu}m}$ with the detector response part about $66.41~\mathrm{{\mu}m}$, which means that the resistive anode readout method can achieve a comparative spatial resolution like the normally used narrow strip readout structure. Similarly, the spatial resolution of other parts in Fig.~\ref{fig_Imaging_Slot} can be calculated as shown in Table~\ref{tab_SpatialResolution}.
\begin{center}
    \includegraphics[width=8cm]{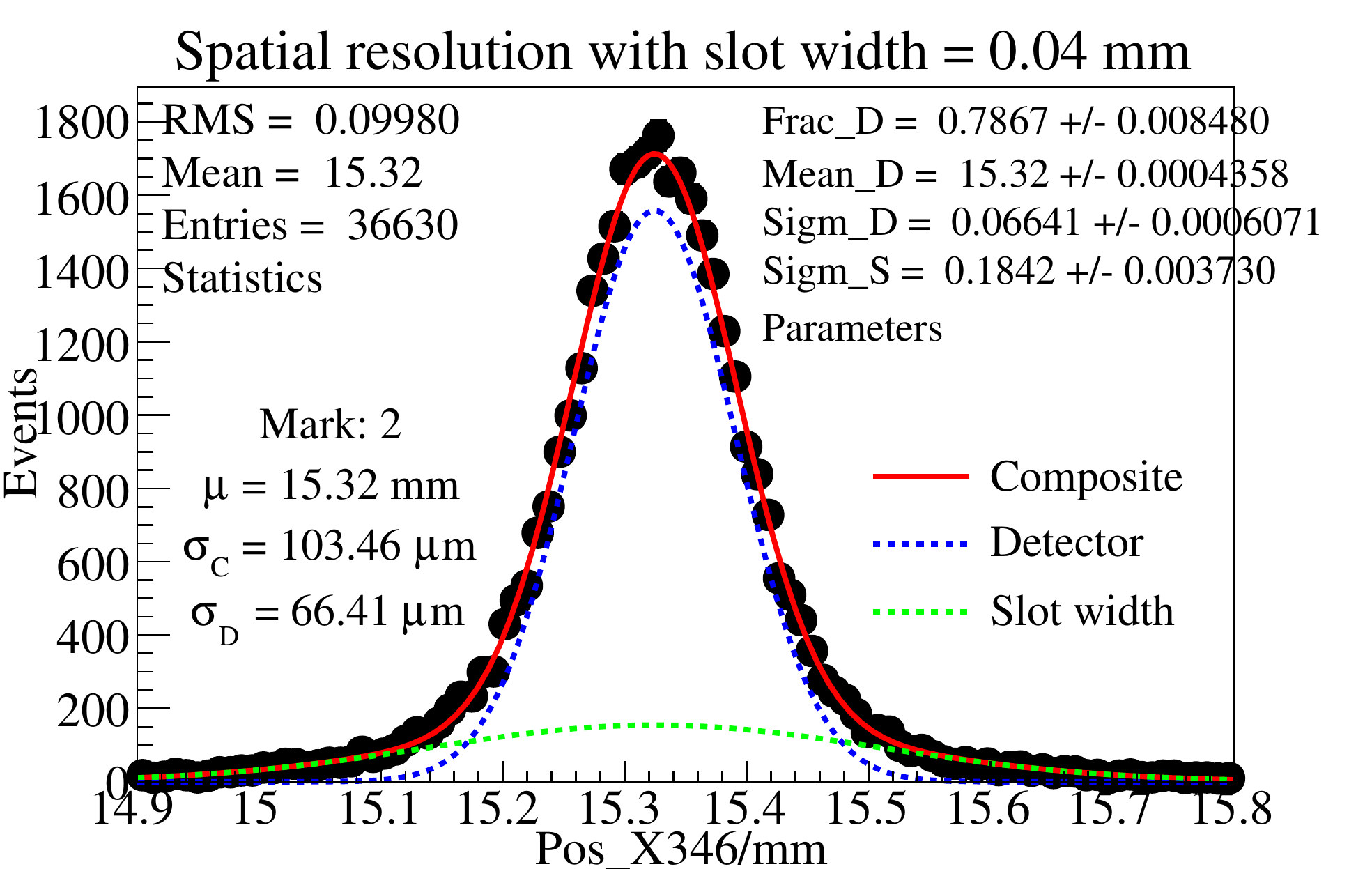}
    \figcaption{\label{fig_SR_Slot-40_Mark2} (color online) Spatial resolution of the detector using the resistive anode readout method. Results were obtained from the mark 2 part data of the imaging of the 40-${\mu}$m-width steel slot in Fig.~\ref{fig_Imaging_Slot} by using the X-ray tube (8~keV).}
\end{center}
\begin{center}
    \tabcaption{ \label{tab_SpatialResolution}  Spatial resolution of different parts in Fig.~\ref{fig_Imaging_Slot}.}
    \footnotesize
    \begin{tabular*}{80mm}{c@{\extracolsep{\fill}}ccc}
        \toprule Mark &${\mu}$~(mm)  &$\sigma({\mu}m)$  &$\sigma_D({\mu}m)$ \\
        \hline
        1  &15.41   &109.39  &\hphantom{0}75.32  \\
        2  &15.32   &103.46  &\hphantom{0}66.41  \\
        3  &15.35   &103.71  &\hphantom{0}66.53  \\        
        4  &15.38   &107.98  &\hphantom{0}73.39  \\        
        \bottomrule
    \end{tabular*}
\end{center}
\par\indent
The difference of the mean values (${\mu}$) in Table~\ref{tab_SpatialResolution} reflects the nonuniformity of the four cells along y axis. Considering the deviation of the surface resistivity~\cite{Juxd-2016} as well as the edge effect of the slot imaging, the nonunoformity of the board, which is less than $0.6\%$, is quite good and leads to a good imaging performance.
\subsection{The nonlinearity}
The linearity of the detector is also an important factor that influences the imaging performance. In order to study the nonlinearity of the detector, a scan along the cells of the resistive anode in x-direction is performed, as shown in Fig.~\ref{fig_Nonlinearity}. 
\begin{center}
    \includegraphics[width=8cm]{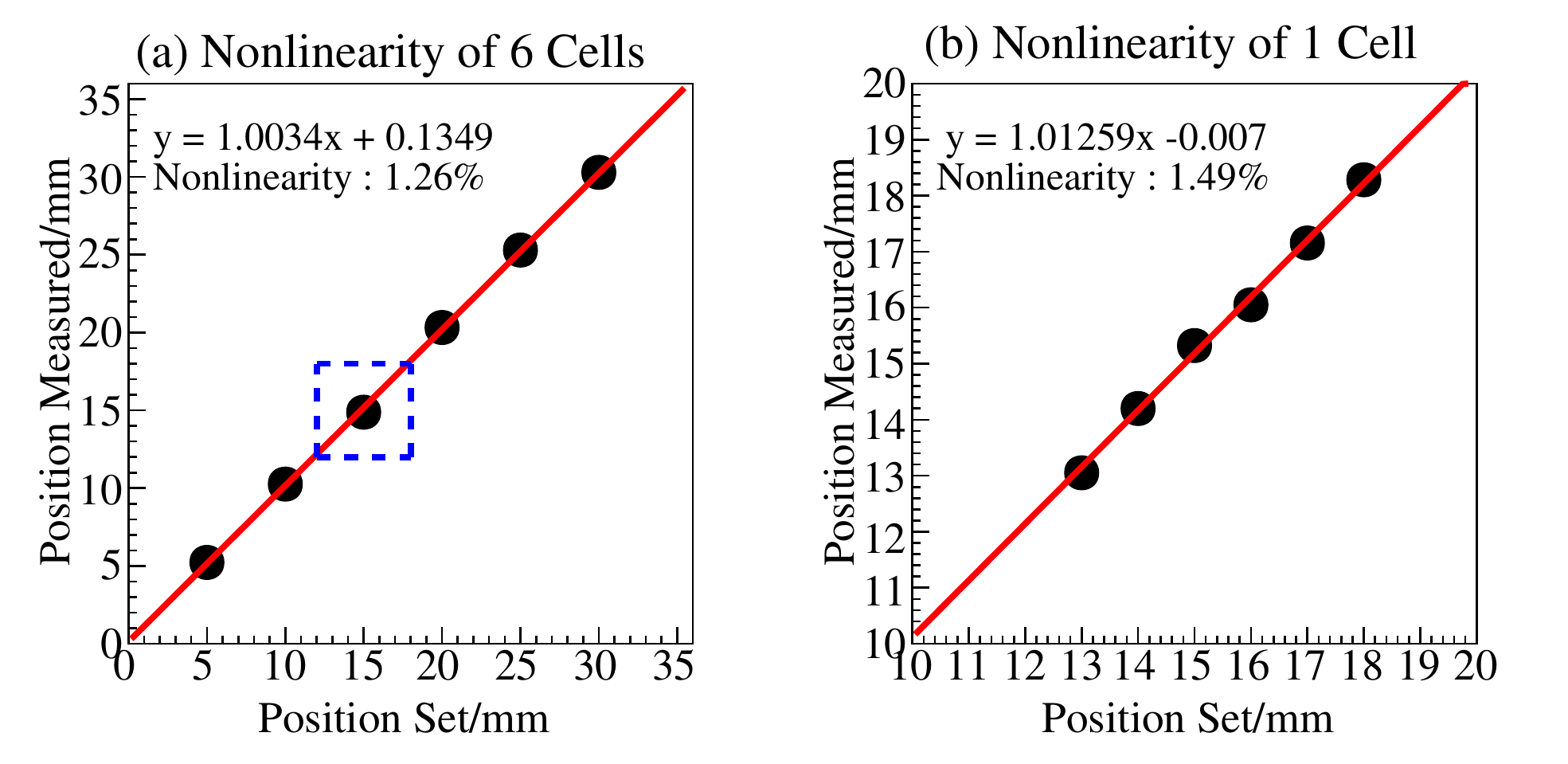}
    \figcaption{\label{fig_Nonlinearity} (color online) Position linearity scan of the resistive anode in x-direction.(a) Scan of all six cells. (b) Scan of one single cell (the dashed blue box in Fig.~\ref{fig_Nonlinearity}-a). Results were obtained from the imaging of a 40-${\mu}$m-width steel slot using an X-ray tube (8~keV). (See in Fig.~\ref{fig_Imaging_Slot})}
\end{center}
\par\indent
The nonlinearity is defined as the average deviation between the fitting positions (based on the set positions) and the measured positions. Fig.~\ref{fig_Nonlinearity}-a shows the scan along the total six cells, and Fig.~\ref{fig_Nonlinearity}-b shows the scan of one single cell. The nonlinearity is less than $1.5\%$, which reflects a good imaging capability.
\subsection{The two dimensional imaging performance}
The main purpose of the GEM detector using the resistive anode readout method is for the two dimensional imaging application area such as the X-ray and neutron imaging. The imaging performance of the detector is tested by using the X-ray tube (8~keV) as well. 
\par\indent
Fig.~\ref{fig_Imaging_key} shows the imaging of a key. The quality of the imaging is quite good. The different shapes of the key edge are clear and correct. The spherical part of the key is also good enough with little distortion. There are defects on the top left corner of the image. The bottom sides of the three cells shrink seriously. The main reason is that the electrical connection between the bottom side of the high resistive pad and the adjacent low resistive strip is bad, which is a common defect found in the experiment for the large area resistive anode.
\begin{center}
    \includegraphics[width=8cm]{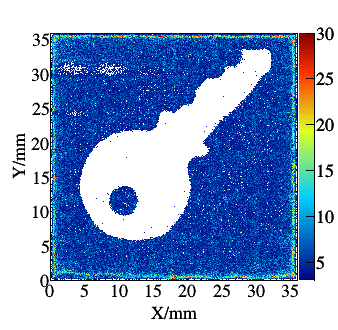}
    \figcaption{\label{fig_Imaging_key} Detector imaging of a key by using an X-ray tube (8~keV).}
\end{center}
\par\indent
Fig.~\ref{fig_Imaging_Pin} shows the imaging of several pin headers with the 2.54~mm pitch. The pins and the insulators separating the pins are discernible clearly. The imaging of parallel pins reflects a good imaging uniformity as well.
\begin{center}
    \includegraphics[width=8cm]{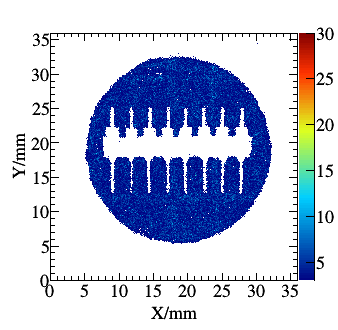}
    \figcaption{\label{fig_Imaging_Pin} Detector imaging of pin headers with the 2.54~mm pitch by using an X-ray tube (8~keV).}
\end{center}
\section{Discussion and summary}
In the paper, we try to apply the resistive anode readout method with more cells for the GEM detector. The resistive anode, which consists of $6\times6$ cells and covers $36\times36~\mathrm{mm}^2$ sensitive area, is manufactured based on the optimisation suggestions mentioned in Ref.~\cite{Juxd-2016}. New electronics and DAQ system are used for the detector, and the signal from the bottom layer of the third GEM foil is chosen as the trigger. The spatial resolution of the detector is $103.46~\mathrm{{\mu}m}$ with the detector response $66.41~\mathrm{{\mu}m}$. What's more, a good two dimensional imaging performance is achieved with good linearity and little distortion. 
\par\indent
In general, the good performances of the detector show the feasibility of the resistive anode readout method with large area for the GEM detector as well as other MPGDs on the two dimensional imgaing application. Furthermore, an integral detector model covering $100\times100~\mathrm{mm}^2$ readout area will be constructed and tested, and the experimental results will be reported later.
\par\indent
\acknowledgments
{
   We are grateful to Dr. Qing-Lei Xiu and Dr. Xin-Yu Lv for the useful discussions.
}
\end{multicols}

\vspace{-1mm}
\centerline{\rule{80mm}{0.1pt}}
\vspace{2mm}

\begin{multicols}{2}

\end{multicols}

\clearpage

\end{CJK*}

\begin{thebibliography}{90}
\vspace{3mm}
\newcommand{\etal}{et al}
\newcommand{\NIMA}{Nucl.\ Instrum.\ Meth.\ A}
\newcommand{\CPC}{Chin.\ Phys.\ C}
\bibitem{Shekhtman-2002}
L.~Shekhtman \etal{}, 
\NIMA{}, 
{\bf 494}: 
128 
(2002)
\bibitem{Sauli-1997}
F.~Sauli \etal{}, 
\NIMA{}, 
{\bf 8}: 
386 
(1997)
\bibitem{Altunbas-2002}
M.~Altunbas \etal{}, 
\NIMA{}, 
{\bf A490}: 
177---203 
(2002)
\bibitem{Thibaud-2014}
F.~Thibaud \etal{}, 
JINST, 
{\bf 9}: 
C02005 
(2014)
\bibitem{Doke-1987}
T.~Doke \etal{}, 
\NIMA{}, 
{\bf 261}: 
605---609 
(1987)
\bibitem{Banu-2008}
A.~Banu \etal{}, 
\NIMA{}, 
{\bf 593}: 
399 
(2008)
\bibitem{Lampton-1979}
M.~Lampton \etal{}, 
Review of Scientific Instruments, 
{\bf 50(9)}: 
1093---1097 
(1979)
\bibitem{Sarvestani-1998}
A.~Sarvestani \etal{}, 
\NIMA{}, 
{\bf 419}: 
444 
(1998)
\bibitem{Orthen-2002}
A.~Orthen \etal{}, 
\NIMA{}, 
{\bf 478}: 
200---204 
(2002)
\bibitem{Wagner-2002}
H.~Wagner \etal{}, 
\NIMA{}, 
{\bf 482}: 
334---346 
(2002)
\bibitem{Wagner-2007}
H.~Wagner \etal{}, 
\NIMA{}, 
{\bf 523}: 
287 
(2004)
\bibitem{Dongmy-2013}
M.~Y.~Dong \etal{}, 
\CPC{}, 
{\bf 37(2)}: 
026002 
(2013)
\bibitem{Xiuql-2013}
Q.~L.~Xiu \etal{}, 
\CPC{}, 
{\bf 37}: 
106002 
(2013)
\bibitem{Juxd-2016}
X.~D.~Ju \etal{}, 
\CPC{}, 
{\bf 40}: 
86004 
(2016)
\bibitem{Guedes-2003}
G.~Guedes \etal{}, 
\NIMA{}, 
{\bf A513}: 
473---483 
(2003)
\bibitem{Lvxy-2012}
X.~Y.~Lv \etal{}, 
\CPC{}, 
{\bf 36}: 
228-234 
(2012)
\bibitem{Zhuys-2006}
Y.~S.~Zhu, 
\textit{Probability and statistics in experimental physics} (Second edition, Beijing:The Science Publishing Company, 2006), P.148
\bibitem{Roofit-2006}
http://roofit.sourceforge.net/docs/RooFit$\_$Users$\_$Manual$\_$2.07-29.pdf, received 11th Jan 2006
\bibitem{Wulh-2007}
L.~H.~Wu, 
\textit{Study of the offline calibration for the BES\uppercase\expandafter{\romannumeral3} drift chamber and the beam test of a prototype}, Ph.D. Thesis(Beijing:Institute of High Energy physics, CAS, 2007)(in Chinese)
\bibitem{Juxd-PhD-2016}
X.~D.~Ju, 
\textit{Study of the Resistive Anode Readout Method Based on the GEM Detector}, Ph.D. Thesis(Beijing:Institute of High Energy physics, CAS, 2016)(in Chinese)
\end{thebibliography}
\end{document}